# ASSESSING SURROGATE ENDPOINTS IN VACCINE TRIALS WITH CASE-COHORT SAMPLING AND THE COX MODEL[1]

By Li Qin, Peter B. Gilbert, Dean Follmann and Dongfeng Li

*Fred Hutchinson Cancer Research Center, Fred Hutchinson Cancer Research Center, National Institute of Allergy and Infectious Diseases and Peking University*

Assessing immune responses to study vaccines as surrogates of protection plays a central role in vaccine clinical trials. Motivated by three ongoing or pending HIV vaccine efficacy trials, we consider such surrogate endpoint assessment in a randomized placebo-controlled trial with case-cohort sampling of immune responses and a time to event endpoint. Based on the principal surrogate definition under the principal stratification framework proposed by Frangakis and Rubin [*Biometrics* **58** (2002) 21–29] and adapted by Gilbert and Hudgens (2006), we introduce estimands that measure the value of an immune response as a surrogate of protection in the context of the Cox proportional hazards model. The estimands are not identified because the immune response to vaccine is not measured in placebo recipients. We formulate the problem as a Cox model with missing covariates, and employ novel trial designs for predicting the missing immune responses and thereby identifying the estimands. The first design utilizes information from baseline predictors of the immune response, and bridges their relationship in the vaccine recipients to the placebo recipients. The second design provides a validation set for the unmeasured immune responses of uninfected placebo recipients by immunizing them with the study vaccine after trial closeout. A maximum estimated likelihood approach is proposed for estimation of the parameters. Simulated data examples are given to evaluate the proposed designs and study their properties.

**1. Introduction.** The evaluation of vaccine efficacy in vaccine clinical trials is generally costly, either because it takes a long trial period for the clinical outcomes to be observed, or because the vaccine may only be partially

Received November 2006; revised August 2007.

[1]Supported by US NIH-NIAID Grant 2 RO1 AI054165-04 and NIH Grant R37 AI29168.

effective. Therefore, identifying vaccine-induced immune responses as surrogate markers for the true study endpoint has spawned interest in vaccine research [Halloran (1998), Chan, Wang and Heyse (2003) and Gilbert et al. (2005)]. The potential surrogate would usually be measured shortly after administration of the study vaccine, and if it can be validated then the vaccine's protective effect can be inferred from it. As knowledge builds on the immunological mechanism for protecting against disease by a pathogen, finding a good immunological surrogate is promising for iteratively guiding refinement of the vaccine formulation, and ultimately for providing a basis for regulatory decisions.

There is an extensive literature on the evaluation of surrogate endpoints for therapeutic development [e.g., Prentice (1989), Lin, Fleming and De Gruttola (1997), DeGruttola et al. (2002), Molenberghs et al. (2002) and Weir and Walley (2006)]. The assessment of an immunological surrogate focuses on contrasting the clinical outcome rate between vaccine recipients and placebo recipients, given the measured immune responses. Since immune response measurements are made post-randomization, this assessment is subject to selection bias [Frangakis and Rubin (2002) and Gilbert, Bosch and Hudgens (2003)]. To address this problem, Gilbert and Hudgens (2006) (henceforth GH) proposed to evaluate the value of a biomarker as a surrogate endpoint by estimating the causal effect predictiveness (CEP) surface, which contrasts the clinical outcome rates between the vaccine recipients and placebo recipients within principal strata formed by joint values of the potential immune responses under assignment to vaccine or placebo. This work built on Frangakis and Rubin (2002)'s potential outcomes framework for evaluating principal surrogate endpoints. GH considered a binary clinical outcome and used a baseline predictor approach to predict the principal strata and estimate the CEP surface nonparametrically. We develop a similar method for a time-to-event clinical endpoint, which is most commonly used in vaccine clinical trials, and use the Cox proportional hazards model [Cox (1972)] to describe the relationship between the survival outcome and covariates including the potential surrogate. Our likelihood calculations utilize discrete failure time models, which are suitable for many vaccine trials because clinical endpoints are often assessed at pre-specified dates.

In the principal stratification framework, the principal strata are subject to missingness as only the immune response to the actual treatment assignment (vaccine or placebo) is observed. This situation was described as the "fundamental challenge of causal inference" [Holland (1986)]. The unobserved immune response is missing for the subjects that receive the "opposite" assignment. We focus on a marginal estimand that conditions on the immune response to the vaccine. Consequently, the assessment of a surrogate in the Cox model framework can be cast as a problem of estimation with a missing covariate. Although methods for estimating the Cox model with missing covariates have been extensively studied [e.g., Lin and Ying (1993), Robins, Rotnitzky and Zhao (1994), Zhou and Pepe (1995), Paik and Tsai (1997), Chen and Little (1999), Herring and Ibrahim (2001), Chen (2002) and Little and Rubin (2002)], their application to the proposed surrogate assessment are not direct, as the missing data are entirely in the placebo group. Techniques are called for to predict the "missing" immune responses in the placebo recipients, or a random sample of them. Therefore, we extend the innovative designs proposed by Follmann (2006) for a binary endpoint to the Cox model setting.

Follmann (2006) proposed two novel components to vaccine trials: baseline irrelevant predictor (BIP), and closeout placebo vaccination (CPV), which enable inference about the vaccine-specific immune responses of placebo recipients. BIP utilizes association between the response of interest and another baseline immune response thought to be irrelevant to infection in the vaccinated subjects. CPV involves vaccinating uninfected placebo recipients after study completion. To match ongoing and pending HIV vaccine trials, we extend these strategies to accommodate a time to event clinical endpoint and sampling of immune responses via a case-cohort design [e.g., Prentice (1986), Borgan et al. (2000), Scheike and Martinussen (2004) and Kulich and Lin (2004)]. We focus on a sampling design that uses data from all infected subjects and a random subcohort of uninfected subjects for whom the immune response to the vaccine is measured (termed "immunogenicity subcohort," $IC$). The methods also apply for other sampling designs, such as failure status-independent case-cohort sampling. We also consider measuring the BIP on some subjects outside the $IC$, which can help improve efficiency.

Under the BIP design placebo subjects cannot be selected into the $IC$; similarly, infected placebo subjects cannot enter $IC$ in the CPV design. Such null selection probabilities violate a key assumption for most semiparametric approaches to handling missing covariates in Cox regression, including all that are based on partial likelihood. Accordingly, we employ a full-likelihood based estimation procedure based on DFT models. For continuous failure time data, we also consider an approximate semiparametric algorithm for the estimation of the BIP-alone design by extending the EM algorithm of Chen (2002).

The proposed methods will be applied to analyze three U.S. National Institutes of Health-sponsored HIV vaccine efficacy trials. These trials randomize HIV negative high risk volunteers to vaccine or placebo in a 1:1 ratio, and follow participants until a fixed number of HIV infection events. The first two trials (named STEP 502 [Mehrotra, Li and Gilbert (2006)] and HVTN 503) are ongoing in the Americas and South Africa, respectively, and evaluate Merck's Adenovirus serotype 5 (Ad5) vector vaccine in approximately 3000 subjects. The third trial (named PAVE-100), co-sponsored by the U.S.

Military HIV Research Program, the International AIDS Vaccine Initiative, and the Centers for Disease Control and Prevention, is being planned. The current PAVE-100 design will randomize approximately 8500 volunteers from 13 countries in the Americas, East Africa, and Southern Africa to placebo or the Vaccine Research Center's prime-boost vaccine regimen (DNA prime:Ad5 vector boost). The trials plan to analyze approximately 100, 120 and 280 HIV infection events, respectively. A secondary objective of each trial is to evaluate the magnitude of $CD8^+$ T cell response levels, as measured by the ELISpot assay from blood samples drawn after Ad5 immunization, as a surrogate for HIV infection. The neutralizing antibody titer to Ad5 is measured at baseline for all participants. Because it is inversely correlated with the $CD8^+$ T cell responses [Catanzaro et al. (2006)], it potentially may be used as a BIP.

To develop our approach for assessing surrogate endpoints in vaccine trials, we present the general framework, assumptions, and definition of the estimands in Section 2, design considerations in Section 3, and an estimation procedure in Section 4. In Section 5 we evaluate the approach with simulated trials designed to match the aforementioned HIV trials. A discussion follows in Section 6.

**2. The principal stratification framework.** In this section we introduce the principal stratification framework based on potential outcomes and principal stratification [Frangakis and Rubin (2002) and Rubin (2005)].

Let $n$ denote the total number of subjects in the vaccine trial. For subject $i$ $(i = 1, \ldots, n)$, let $V_i$ denote the observed treatment indicator, $W_i$ denote a collection of first-phase baseline covariates in the case-cohort sampling (measured on everyone), and $S_i(V)$ denote the potential immune response of the subject if he/she is assigned vaccine $(V = 1)$ or placebo $(V = 0)$. Similarly, for $V = 1, 0$, let $T_i(V)$ and $C_i(V)$ be the potential failure time and censoring time, and $X_i(V) = \min\{T_i(V), C_i(V)\}$ and $\delta_i(V) = I(T_i(V) \leq C_i(V))$. Let $t_1, \ldots, t_K$ indicate the fixed visit times, with $t_2, \ldots, t_K$ the possible discrete failure times for $X_i(V_i)$. Let $t_K^+$ denote censored at the final visit and $M_i$ denote the last visit number of subject $i$ during the trial period, thus, $M_i \in \{1, \ldots, K\}$. For vaccine recipients at-risk at $t_1$ and in the $IC$, the immune response $S_i(V)$ is measured at time $t_1$. Letting $R_i(V)$ denote the potential at-risk indicator at $t_1$, $S_i(V)$ is only defined if $R_i(V) = 1$; otherwise, we put $S_i(V) = *$. We assume that the censoring process $C_i(V)$ and failure time distribution $T_i(V)$ are independent given $\{W_i, R_i(V), S_i(V)\}$.

Suppose that $\{V_i, W_i, R_i(0), R_i(1), S_i(0), S_i(1), X_i(0), X_i(1), \delta_i(0), \delta_i(1), i = 1, \ldots, n\}$ are i.i.d. We make the following assumptions to identify the estimands:

A1. Stable unit treatment value assumption (SUTVA).

A2. *Ignorable treatment assignments.* Conditional on $W_i$, $V_i$ is independent of $\{R_i(0), R_i(1), S_i(0), S_i(1), X_i(0), X_i(1), \delta_i(0), \delta_i(1)\}$.

Assumption A1 guarantees the "consistency" property (i.e., the observed outcomes for a subject assigned $V$ equals his potential outcomes if assigned $V$) and that the potential outcomes of one subject are not impacted by the treatment assignments of other subjects. A2 holds for randomized, blinded trials.

Under the above assumptions, we define two vaccine efficacy estimands:

1. *Conditional on joint potential outcomes* (*joint VE*)

$$VE(s_1, s_0)$$
$$\equiv 1 - \frac{\Pr(T(1) = t_k | T(1) \geq t_{k-1}, S(1) = s_1, S(0) = s_0, R(1) = 1, R(0) = 1)}{\Pr(T(0) = t_k | T(0) \geq t_{k-1}, S(1) = s_1, S(0) = s_0, R(1) = 1, R(0) = 1)}.$$

2. *Conditional on marginal potential outcome* (*marginal VE*)

$$VE(s_1) \equiv 1 - \frac{\Pr(T(1) = t_k | T(1) \geq t_{k-1}, S(1) = s_1, R(1) = 1)}{\Pr(T(0) = t_k | T(0) \geq t_{k-1}, S(1) = s_1, R(1) = 1)},$$
$$k = 2, \ldots, K.$$

The estimand $VE(s_1, s_0)$ conditions on membership in the basic principal stratum $\{S(1) = s_1, S(0) = s_0, R(1) = R(0) = 1\}$, and $VE(s_1)$ conditions on membership in a union of basic principal strata [Frangakis and Rubin (2002)]. The estimands condition on $R_i(1) = R_i(0) = 1$ or on $R_i(1) = 1$ because $S_i(V)$ is only defined if $R_i(V) = 1$, $V = 0, 1$. The estimands are principal stratification estimands in that the pair $(S(1), S(0))$ or $S(1)$ can be treated as a baseline covariate. However, they are not causal estimands, because the numerators and denominators condition on different events $T(1) \geq t_{k-1}$ and $T(0) \geq t_{k-1}$. Nevertheless they are scientifically interesting, in the same way that a hazard ratio conditional on baseline covariates is interesting.

To help identify the estimands, only subjects with $R_i(V_i) = 1$ are included in the analysis, and we assume the following:

A3. *Equal drop-out and risk up to time* $t_1$: $R_i(1) = 1 \Longleftrightarrow R_i(0) = 1$.

A3 implies that subjects observed to be at risk at $t_1$ will have $R_i(1) = R_i(0) = 1$, so that $S_i(1)$ and $S_i(0)$ are both defined.

In addition to A1–A3, identifiability of $VE(s_1, s_0)$ requires a way to predict $S_i(1)$ for subjects with $V_i = 0$ and a way to predict $S_i(0)$ for subjects with $V_i = 1$. Identifiability of $VE(s_1)$ is easier because only the $S_i(1)$ for subjects in arm $V_i = 0$ must be predicted. Furthermore, for our motivating application, typically the immune response $S_i(0)$ is zero for all placebo recipients, because exposure to the vaccine is necessary to stimulate an immune response. For these reasons, henceforth, we focus on the marginal estimand $VE(s_1)$. Note that, for applications with $S_i(0) = 0$ for all $i$, $VE(s_1) = VE(s_1, 0)$.

We propose a Cox model for the discrete cumulative hazard function $\Lambda(t)$,

$$d\Lambda(t_k; V, S(1) = s_1, R(1) = 1, W) = \exp(Z'\boldsymbol{\beta})d\Lambda_0(t_k),$$

(1)

$$k = 2, \ldots, K,$$

with $Z = \{V, S(1), VS(1), W'\}'$, $\boldsymbol{\beta} = \{\beta_1, \beta_2, \beta_3, \boldsymbol{\beta}_4'\}'$, and $\Lambda_0(\cdot)$ is the discrete baseline cumulative hazard function. The marginal $VE(s_1)$ can be expressed as

$$VE(s_1) = 1 - \frac{d\Lambda(t_k; V = 1, S(1) = s_1, R(1) = 1)}{d\Lambda(t_k; V = 0, S(1) = s_1, R(1) = 1)}, \qquad k = 2, \ldots, K.$$

The discrete hazards always condition on $\{R(1) = 1\}$ and, henceforth, we assume this implicitly. For subjects with a particular baseline covariate $w$, a similar estimand $VE(s_1|w)$ can be formed by conditioning on $W = w$ in the hazards.

The population estimand $VE(s_1)$ contrasts the rate of the clinical event for subjects with $S(1) = s_1$ under assignment to vaccine versus under assignment to placebo. Supposing $S(1)$ is bounded below at value zero which indicates a negative immune response, we define $S$ to be a *predictive surrogate* if $VE(0) = 0$ and $VE(s_1) > 0$ for all $s_1 > C$ for some constant $C \geq 0$. These conditions reflect population level necessity and sufficiency of the immune response to achieve positive vaccine efficacy.

Under A1–A3 and the Cox model (1), the estimand equals

(2)                     $$VE(s_1) = 1 - \exp(\beta_1 + s_1\beta_3).$$

In equation (2) a negative value of $\beta_3$ indicates that a higher immune response to vaccine predicts greater vaccine efficacy. On the other hand, $\beta_3 = 0$ implies $VE(s_1)$ is constant in $s_1$ so that the marker does not predict vaccine efficacy. Therefore, testing $H_0 : \beta_3 = 0$ versus $H_1 : \beta_3 < 0$ assesses sufficiency. A value $\beta_1 = 0$ indicates necessity, and both $\beta_1 = 0$ and $\beta_3 < 0$ indicate the marker is a predictive surrogate. The magnitude of $\beta_3$ indicates the quality of the predictive surrogate with $\beta_3 = 0$ suggesting no surrogate value [$VE(s_1)$ is constant in $s_1$] and larger $|\beta_3|$ suggesting greater surrogate value (greater predictiveness).

**3. Augmented designs for estimation.** The immune response to the study vaccine, $S(1)$, cannot be measured in placebo recipients, but it may be inferred when utilizing either the BIP or CPV designs (see Figure 1).

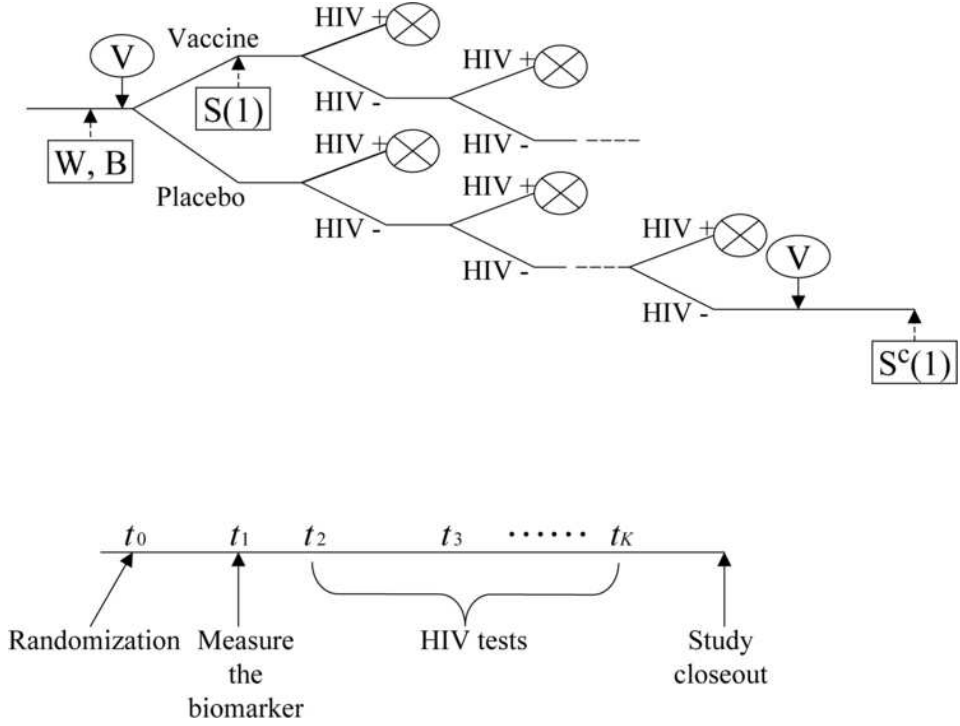

FIG. 1. *Illustration of an HIV vaccine trial design under the BIP and CPV strategies. Under BIP or BIP+CPV, baseline measurements of W and B are obtained from all (or a random sample of) the study participants prior to the randomization at time 0. The study subjects are then randomized to receive inoculation V of the study vaccine or placebo. For some vaccine recipients, the immune response to the vaccine $S(1)$ is measured at time $t_1$. The subsequent assessments of HIV infection are conducted at discrete times $t_2, \ldots, t_K$. The study subjects are followed until diagnosis of HIV infection (HIV+) or study closeout at or after $t_K$. Under CPV or BIP+CPV, placebo recipients uninfected (HIV−) at study closeout (or a random sample of them) are immunized with the study vaccine and the immune response $S^c(1)$ is measured $t_1$ units of time afterward.*

*Baseline Irrelevant Predictor (BIP).* Assume a baseline covariate $B$ is available that does not affect (i.e., is "irrelevant" for) clinical risk after accounting for the immune response $S(1)$ and first-phase covariates $W$:

A4. $d\Lambda(t_k; V, S(1), W, B) = d\Lambda(t_k; V, S(1), W), \ k = 2, \ldots, K, \ V = 0, 1.$

Assumptions A1–A3 imply that the relationship between $S(1)$ and $B$ is the same regardless of treatment assignment

$$(3) \qquad [S_i(1)|V_i = 1, B_i, R_i(1) = 1] \overset{d}{=} [S_i(1)|V_i = 0, B_i, R_i(1) = 1].$$

Therefore, $S_i(1)$ can be predicted or imputed for placebo subjects based on $B_i$. For vaccine recipients with the BIP measured and who are outside the $IC$, their immune responses are predicted using the BIP as well.

In case-cohort designs, good baseline predictors need to be highly correlated with the biomarker $S(1)$, and preferably include first-phase (measured on everyone) inexpensive covariates to achieve efficiency gains.

*Closeout Placebo Vaccination* ($CPV$). This design entails vaccinating uninfected placebo subjects after the study closeout, and measuring their immune response $S_i^c(1)$. The closeout measurement $S_i^c(1)$ is made at a visit $t_1$ time units after vaccination, to match the measurement schedule in the vaccine trial. We need to make an additional assumption to bridge the marker values $S_i(1)$ and $S_i^c(1)$. Let $S_i^{\text{true}}(1)$ be the true immune response at time $t_1$, allowing that the observed immune response is subject to some assay measurement error.

A5. *Time constancy of $S_i^{\text{true}}(1)$*: For uninfected placebo recipients, $S_i(1) = S_i^{\text{true}}(1) + e_{i1}$ and $S_i^c(1) = S_i^{\text{true}}(1) + e_{i2}$, where $e_{i1}$ and $e_{i2}$ are independent and identically distributed random errors with mean 0.

This assumption implies that the true immune response is unchanged from time $t_1$ to study closeout plus $t_1$, and the measurement errors have the same distribution. Thus, $S_i(1)$ and $S_i^c(1)$ are exchangeable and one can be used in lieu of the other. To be concrete, suppose only one shot is given, the trial is three years, and $t_1$ is 6 months after the shot. A5 states that the true immune response 6 months after the shot is the same whether it is measured January 1, 2004 or January 1, 2007. In the Discussion we outline how our methods can be generalized to use $S_i^{\text{true}}(1)$ in the Cox model (1) rather than $S_i(1)$. Note that even if the regression involves $S_i^{\text{true}}(1)$, a valid test of the effect of $S_i^{\text{true}}(1)$ obtains when using $S_i(1)$ [Prentice (1982)]. If time constancy of immune response is not reasonable, then $S_i^c(1)$ cannot be used in lieu of $S_i(1)$ and CPV may be questionable. See Follmann (2006) for further discussion of this issue, including how to examine this assumption.

Under A5, the distribution of $[S_i(1)|V_i = 0, \delta_i = 1]$ can be inferred from the marginal distributions $[S_i(1)|V_i = 1] \stackrel{d}{=} [S_i(1)|V_i = 0]$. However, in case-cohort sampling, if the $IC$ is small, then the large amount of missing data and the inferred immune responses in placebo recipients may challenge the performance of the method.

*Baseline irrelevant predictor and closeout placebo vaccination combined* ($BIP + CPV$). The BIP and CPV designs can be combined by imputing $S_i(1)$ with $S_i^c(1)$ for all uninfected placebo recipients with $S_i^c(1)$ measured, and predicting $S_i(1)$ with $B_i$ for all others with $B_i$ measured. Combining the designs can yield large efficiency gains. In the situation where there is no good baseline predictor or the baseline predictor is expensive to collect, conducting small-scale CPV on a random sample of the uninfected placebo recipients can add accuracy and precision to the estimates.

**4. Estimation.** Estimation of the estimand is challenged by the amount of missing $S(1)$'s. We focus on the maximum estimated likelihood (MEL) estimation procedure that applies to all three designs. We then briefly outline an approximate EM-type algorithm for estimation with the BIP-alone design.

4.1. *Maximum estimated likelihood estimation.* We present below the estimation procedure for the BIP + CPV design, which includes estimation under the BIP- or CPV-alone designs as special cases.

Let $IC_V$ denote the immunogenicity cohort that contributes second-phase data $S(1)$ in vaccine recipients, and $IC_P$ denote the cohort within uninfected placebo subjects that received vaccination at study closeout, so that $IC = IC_V \cup IC_P$. Let $IB$ denote the set of subjects with $B$ measured, which can be larger than $IC$. For placebo subjects that do not have $S(1)$ measured, their likelihood contribution integrates over the marginal distribution of $S(1)$ or the conditional distribution of $S(1)|B$. The full log-likelihood of model (1) under the BIP + CPV design (with convention that $\prod_{j=2}^{1} = 1$) is given by

$$\log L(\boldsymbol{\beta}, \boldsymbol{\lambda}_0) = \sum_{i \in IC_V} \log L_1(O_i) + \sum_{i \in IC_P} \log L_2(O_i) + \sum_{i \in \overline{IC}, IB} \log L_3(O_i)$$
$$(4) \qquad \qquad + \sum_{i \in \overline{IC}, \overline{IB}} \log L_4(O_i),$$

where

$$L_1(O_i) = \prod_{j=2}^{M_i - 1} (1 - \lambda_{0j})^{\exp\{V_i\beta_1 + S_i(1)\beta_2 + V_i S_i(1)\beta_3 + W_i'\boldsymbol{\beta}_4\}R_i(V_i)}$$

$$\times \left\{ 1 - (1 - \lambda_{0,M_i})^{\exp\{V_i\beta_1 + S_i(1)\beta_2 + V_i S_i(1)\beta_3 + W_i'\boldsymbol{\beta}_4\}} \right\}^{\delta_i R_i(V_i)}$$

$$\times (1 - \lambda_{0,M_i})^{\exp\{V_i\beta_1 + S_i(1)\beta_2 + V_i S_i(1)\beta_3 + W_i'\boldsymbol{\beta}_4\}(1 - \delta_i)R_i(V_i)},$$

$$L_2(O_i) = \prod_{j=2}^{M_i} (1 - \lambda_{0j})^{\exp\{V_i\beta_1 + S_i^c\beta_2 + V_i S_i^c\beta_3 + W_i'\boldsymbol{\beta}_4\}R_i(V_i)},$$

$$L_3(O_i) = \int \prod_{j=2}^{M_i - 1} (1 - \lambda_{0j})^{\exp\{V_i\beta_1 + s\beta_2 + V_i s\beta_3 + W_i'\boldsymbol{\beta}_4\}R_i(V_i)}$$

$$\times \left\{ 1 - (1 - \lambda_{0,M_i})^{\exp\{V_i\beta_1 + s\beta_2 + V_i s\beta_3 + W_i'\boldsymbol{\beta}_4\}} \right\}^{\delta_i R_i(V_i)}$$

$$\times (1 - \lambda_{0,M_i})^{\exp\{V_i\beta_1 + s\beta_2 + V_is\beta_3 + W_i'\boldsymbol{\beta}_4\}(1-\delta_i)R_i(V_i)} \, dP(s|B_i, W_i),$$

$$L_4(O_i) = \int \prod_{j=2}^{M_i-1} (1 - \lambda_{0j})^{\exp\{V_i\beta_1 + s\beta_2 + V_is\beta_3 + W_i'\boldsymbol{\beta}_4\}R_i(V_i)}$$

$$\times \{1 - (1 - \lambda_{0,M_i})^{\exp\{V_i\beta_1 + s\beta_2 + V_is\beta_3 + W_i'\boldsymbol{\beta}_4\}}\}^{\delta_i R_i(V_i)}$$

$$\times (1 - \lambda_{0,M_i})^{\exp\{V_i\beta_1 + s\beta_2 + V_is\beta_3 + W_i'\boldsymbol{\beta}_4\}(1-\delta_i)R_i(V_i)} \, dP(s|W_i).$$

Here $\boldsymbol{\lambda}_0 = \{\lambda_{02}, \ldots, \lambda_{0K}\}^T$ are unknown baseline hazards (with $\lambda_{0k} = d\Lambda_0(t_k)$, $k = 2, \ldots, K$), and $P(s|w)$ and $P(s|b, w)$ are the conditional c.d.f.'s of $S(1)$.

In the Cox model formulation, the estimand $VE(s_1)$ depends only on $\boldsymbol{\beta}$ while the parameters in the conditional c.d.f.'s $P(s|w)$ and $P(s|b, w)$ are nuisance parameters. Rather than maximizing the full likelihood over the entire parameter space, we take the MEL approach [Pepe and Fleming (1991)] to avoid specifying the joint distribution of $(S(1), B, W)$ and the intensive computations entailed in the numerical integration. The conditional c.d.f.'s $P(s|w)$ and $P(s|b, w)$ are first consistently estimated from the vaccine recipients' data (Section 4.1.1), and then the estimated likelihood $\log L(\boldsymbol{\beta}, \boldsymbol{\lambda}, \widehat{P}(\cdot), \widehat{P}(\cdot|\cdot))$ is constructed.

For a categorical $W$, $P(s|w)$ and $P(s|b, w)$ can be estimated nonparametrically. However, if $W$ is continuous, then nonparametric estimation will require smoothing and much larger sample sizes are needed for tractable computation. Therefore, if $W$ is continuous or multi-component, parametric assumptions on the conditional c.d.f.'s will usually be needed to achieve stable estimation in practice. An advantage of the MEL approach is that it can straightforwardly accommodate any approach to estimating the nuisance parameters $P(s|w)$ and $P(s|b, w)$. In the MEL approach we first estimate these distributions consistently using data from the vaccine recipients, and then construct the estimated likelihood $L(\boldsymbol{\beta}, \boldsymbol{\lambda}, \widehat{P}(\cdot), \widehat{P}(\cdot|\cdot))$.

We outline three key steps in the evaluation of the log-likelihood (4) in the absence of the first-phase covariates $W$:

1. *Estimation of $p(s)$ and $p(s|b)$.* Let $p(s)$, $p(b)$, and $p(s, b)$ be marginal and joint p.d.f.s (or p.m.f.s for discrete variables) for $S(1)$ and $B$. Because vaccine recipients in the $IC_V$ provide nonrandom samples of $S(1)$ and $B$, and vaccine recipients in the $IB$ contribute additional data for $B$, it follows that

$$
\begin{aligned}
p(s) &= f_{11}(s)p_{11} + f_{10}(s)p_{10}, \\
p(b) &= f_{11}(b)p_{11} + f_{10}(b)p_{10}, \\
p(s, b) &= f_{11}(s, b)p_{11} + f_{10}(s, b)p_{10},
\end{aligned}
$$

(5)

where, for $h = 1, 0$, $f_{1h}(\cdot)$ is the conditional p.d.f. or p.m.f. of $S(1)$ given $V = 1$ and $\delta = h$, and $p_{1h} \equiv \Pr(\delta = h|V = 1)$. The probabilities $\{p_{1h}\}$ can be estimated by their sample counterparts $\{\hat{p}_{1h}\}$ and estimates of $\{f_{1h}(s), f_{1h}(b), f_{1h}(s, b)\}$.

We sketch the estimation for two special cases where (A) $(S(1), B)$ are categorical and (B) $(S(1), B)$ are bivariate normally distributed.

(A) If $S(1)$ and $B$ have discrete values with $J$ and $L$ categories, respectively, then $f_{1h}(s_j)$ and $p(S(1) = s_j | b_l)$ $(j = 1, \ldots, J, l = 1, \ldots, L)$ can be estimated nonparametrically:

$$\hat{f}_{1h}(s_j) = \frac{\sum_{i \in IC_V} I(S_i(1) = s_j, \delta_i = h)}{\sum_{i \in IC_V} I(\delta_i = h)},$$

$$\hat{p}(S(1) = s_j | b_l) = \frac{\sum_{i \in IC_V, B_i = b_l} \delta_i I(S_i(1) = s_j)}{\sum_{i \in IC_V, B_i = b_l} \delta_i} \hat{p}_{11}$$

$$+ \frac{\sum_{i \in IC_V, B_i = b_l} (1 - \delta_i) I(S_i(1) = s_j)}{\sum_{i \in IC_V, B_i = b_l} (1 - \delta_i)} \hat{p}_{10}.$$

(B) If $(S(1), B)$ are jointly normally distributed, then $p(s)$ and $p(s|b)$ are both normal densities and thus can be estimated using estimates of the first and second moments from expressions in (5).

Evaluating the likelihood (4) involves integrations over $s$, which are briefly described in the Appendix.

2. *Maximization and implementation.* The estimated log-likelihood $\log L(\boldsymbol{\beta}, \boldsymbol{\lambda}, \hat{P}(\cdot), \hat{P}(\cdot|\cdot))$ is maximized using quasi-Newton methods. The assumption that $S(1)$ is observed with nonzero probability in all subjects is violated. Therefore, the asymptotic variance of $\hat{\boldsymbol{\beta}}$ via the MEL approach cannot be derived analytically. We propose to obtain the standard errors for $\hat{\boldsymbol{\beta}}$ by the bootstrap. For computational efficiency, the software for estimation is implemented in Matlab 7.0.1 (Mathworks, Inc) with a C++ plug in, compiled to dynamic link library.

4.2. *Approximate EM-type estimation.* In this subsection we present an estimation approach that uses regression calibration to impute the missing $S_i(1)$s for subjects with a BIP $B_i$ measured and employs an EM-type algorithm based on full likelihood to accommodate the missing $S_i(1)$s for subjects without $B_i$ measured. Because the CPV-based designs have missing $S(1)$s for the entire $\{V = 0, \delta = 1\}$ stratum, the algorithm can only reliably estimate the Cox model parameters for the BIP-alone design, as confirmed in simulations. We focus on the BIP-alone design with a continuous BIP in this section. The proposed algorithm can be applied to a categorical BIP with slight modification. An advantage of this EM approach is that it accommodates continuous failure times.

Because the missingness of $S(1)$ does not depend on unobserved $S(1)$, and we assume the censoring distribution does not depend on $S(1)$, the log-likelihood for the BIP-alone design can be expressed up to a constant factor

as

$$l(\boldsymbol{\beta}, \boldsymbol{\alpha}, \Lambda_0)$$
$$= \sum_{i \in IC} \{\delta_i(Z_i'\boldsymbol{\beta}) - \Lambda_0(X_i)\exp(Z_i'\boldsymbol{\beta})\}$$
$$+ \sum_{i \in \overline{IC}, IB} \log\left\{\int \exp\{\delta_i(Z_i'\boldsymbol{\beta}) - \Lambda_0(X_i)\exp(Z_i'\boldsymbol{\beta})\} \, dP(s|V_i, W_i, B_i)\right\}$$
$$+ \sum_{i \in \overline{IC}, \overline{IB}} \log\left\{\int \exp\{\delta_i(Z_i'\boldsymbol{\beta}) - \Lambda_0(X_i)\exp(Z_i'\boldsymbol{\beta})\} \, dP(s|V_i, W_i)\right\}$$
$$+ \delta_i \log(d\Lambda_0(X_i)),$$

where $X_i$ denotes the observed failure time, $\Lambda_0(X)$ denotes the baseline cumulative hazard function, and $\boldsymbol{\alpha}$ represents unknown parameters in the conditional distributions of $S(1)$.

The log-likelihood score equations can be solved via an iterative EM algorithm [Chen and Little (1999), Herring and Ibrahim (2001), Chen (2002)]. For computational efficiency, we propose to modify the double-semiparametric EM-algorithm of Chen (2002) to incorporate the auxiliary covariate $B$ as a predictor of the missing $S(1)$. Given equation (3) and the relationship $S_i(1) = g(B_i; \theta) + \epsilon_i$, where $g(\cdot)$ is a parametric link function depending on the unknown parameter $\theta$ and $\epsilon_i$ has mean zero and variance $\sigma^2$, $S_i(1)$ can be predicted by $\widehat{E}(S(1)|B_i) = g(B_i; \widehat{\theta})$. When the event occurrence is rare, $E(S(1)|B_i) \approx E(S(1)|B_i, X_i, \delta_i)$. This fact has been well studied in the context of regression calibration in the Cox regression [e.g., Prentice (1982) and Wang et al. (1997)]. Therefore, unobserved $S(1)$'s can be replaced by $\widehat{E}(S(1)|B)$ and treated as observed data in the EM algorithm. We name this procedure the "Approximate Calibration-Based EM (ACEM)" algorithm. An outline of this procedure is given below; interested readers are referred to Chen (2002) for details:

1. Calibration-step: Prediction of unobserved $S_i(1)$s by $\widehat{S}_i(1) = \widehat{E}(S(1)|B_i)$.
2. E-step: Given parameter values at the $m$th iteration ($\boldsymbol{\beta}^{(m)}, \Lambda_0^{(m)}(X), \boldsymbol{\alpha}^{(m)}, p_{klj}^{(m)}, \theta^{(m)}$), for $p_{klj}$ denote the probability mass of the observed distinct values of $S(1)$ at discrete levels of $V = v_k$ and $W_d = w_l$ ($W = W_d \cup W_c$ where $W_d$ and $W_c$ denote the categorical and continuous covariates in $W$, resp.), and $\boldsymbol{\alpha}^{(m)}$ denote the parameters in the distribution $P(W_c|S(1), V, W_d, X, \delta)$. Calculate conditional expectations under $P(S(1)|V, W_d, X, \delta)$.
3. M-step: Update ($\boldsymbol{\beta}, \Lambda_0(X), \boldsymbol{\alpha}, p_{klj}, \theta$) by solving the corresponding score equations.

4. Repeat the E-step and M-step above until convergence.

The advantage of the ACEM algorithm is that it can account for continuous failure times and is computationally fast; however, since it uses regression calibration, it performs well only for the rare event situation with a highly predictive BIP. Prevention trials, which usually have a low event rate, are an application area.

**5. Simulation study.** We conducted a simulation study to evaluate the performance of the proposed strategies for estimating the estimand $VE(s_1)$ and thereby assessing a predictive surrogate in the Cox model setting. To simulate the real scenarios, we roughly follow the design of the three HIV vaccine efficacy trials described in the introduction. We suppose a total sample size of 5000, with 2500 subjects per arm. The treatment indicator $V = 1$ if assigned vaccine and $V = 0$ if assigned placebo. Under the case-cohort sampling, the immunogenicity subcohort ($IC$) consists of all infected vaccine recipients and a random sample of uninfected vaccine recipients, which include a combination of 25% or 50% of uninfected vaccine recipients. We considered one auxiliary covariate $B$ as the BIP for the potential immunological surrogate $S(1)$. The variables $S(1)$ and $B$ were generated from a bivariate normal distribution with mean zero and variance 0.4 for each component [reflecting the variance of the ELISPOT assay used to measure $S(1) = \mathrm{CD8^+}$ T cell response], and correlation $\rho = 0.6$ or 0.9. For the BIP-alone and BIP + CPV designs, we assume that $B$ was measured from all individuals in the $IC$ and from 50% or 37.5% of those not in the $IC$, as a precision factor. In the BIP-alone approach, $S(1)$ was treated as missing for all placebo recipients, while for the BIP + CPV and CPV-alone approaches, we assume 25% or 50% uninfected placebo recipients got the CPV measurement $S^c(1)$. Infection times were generated from the continuous-time Cox model $\lambda(t|V, S(1)) = \lambda_0(t) \exp\{\beta_1 V + \beta_2 S(1) + \beta_3 V S(1)\}$, and were grouped into 6 equal-length time intervals to reflect the discrete visit schedule of the trials. The true parameters $\beta_2 = -1.109$ and $\beta_3$ were set at 0, $-0.4$, or $-0.7$, reflecting the null hypothesis that $S(1)$ has no value as a predictive surrogate and alternative hypotheses of 1.2-fold and 1.5-fold lower relative risks $RR(S(1)) = 1 - VE(S(1))$ per 1 standard deviation higher immune response $S(1)$, corresponding to low and high surrogate value, respectively. In addition, $\lambda_0(t) = \lambda_0$ and $\beta_1$ were calibrated to give $VE(0) = 0.5$ and 334 infections expected in the placebo arm, and hence, 7% overall infection rate. Random censoring of 10% was added to account for subject drop out. All uninfected subjects were censored at the end of the follow-up period, specified at 3 years. Five hundred simulation runs and 50 bootstrap replicates were used to obtain standard error estimates for the estimated regression parameters.

We first conducted estimation through the MEL algorithm for discrete failure times using all three designs. For the BIP-alone design, a second simulation was conducted to compare the performance of the MEL approach for grouped failure times, versus that of the ACEM algorithm assuming continuous failure times were observed in a rare event setting. To evaluate efficiencies for the parameter estimates, estimates from the Cox model using the full simulated data were obtained as an unattainable "gold standard."

Figure 2 plots the true $VE(s_1)$ curve for different true parameters $(\beta_1, \beta_3)$ in model (2). It shows that when $\beta_3 = -0.7$, $VE(0) = 0$ and $VE(s_1) > 0$ for $s_1 > 0$, indicating that the immune response variable is a predictive surrogate.

Table 1 presents simulation results for the MEL approach in different settings. It can be seen that the method has excellent performance. There are generally small biases, small variances of the estimates and good power of the test of $H_0 : \beta_3 = 0$ for surrogate value. As more CPV or auxiliary BIP information is available, both the accuracy and precision of the estimates improve. The efficiency of the BIP-involved designs increases as the correlation between the BIP and $S(1)$ increases. The CPV-alone design is less

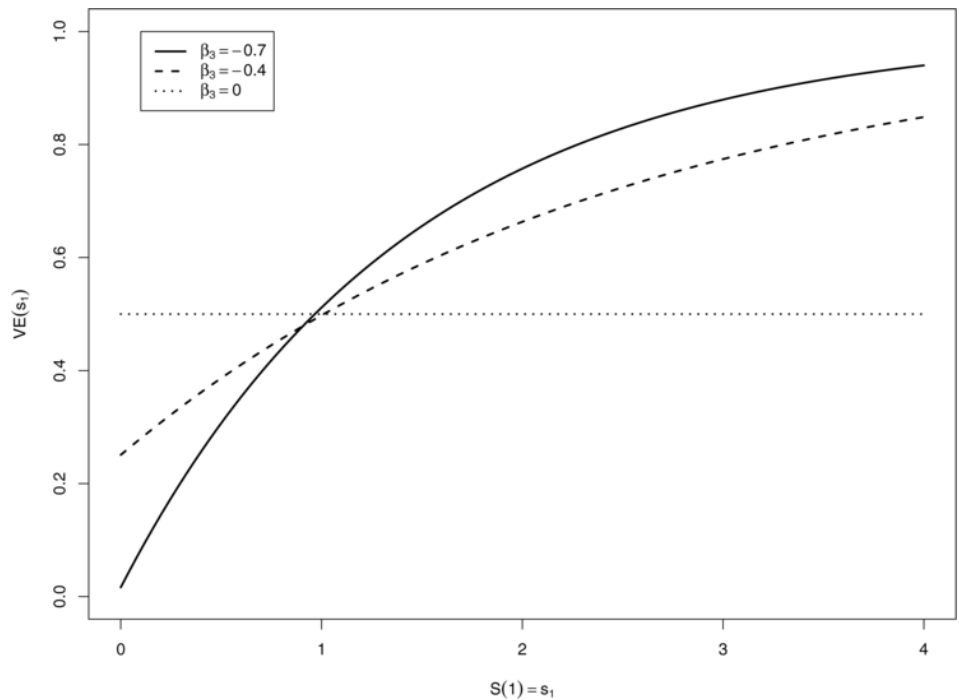

Fig. 2. *Illustration of the estimand $VE(s_1)$ as a function of the standardized potential surrogate $S(1)$ over the range of observable values with different values for $\beta_3$.*

TABLE 1
*Results from the MEL estimation*

| Design | $\rho$ | $\beta$ | Missing | $\hat{\beta}_1$ | | | | $\hat{\beta}_2$ | | | | $\hat{\beta}_3$ | | | | |
|---|---|---|---|---|---|---|---|---|---|---|---|---|---|---|---|---|
| | | | | Bias | SD | SE | RE | Bias | SE | ASE | RE | Bias | SD | SE | RE | Power |
| BIP | 0.6 | $\beta^{(0)}$ | Large | 0.004 | 0.13 | 0.14 | 26 | 0.007 | 0.22 | 0.24 | 9 | −0.013 | 0.28 | 0.29 | 15 | 5 |
| | | | Medium | 0.009 | 0.13 | 0.14 | 26 | −0.001 | 0.22 | 0.22 | 9 | −0.009 | 0.27 | 0.27 | 16 | 5 |
| | | $\beta^{(4)}$ | Large | −0.001 | 0.08 | 0.08 | 84 | 0.007 | 0.18 | 0.18 | 14 | −0.010 | 0.21 | 0.20 | 32 | 52 |
| | | | Medium | 0.001 | 0.08 | 0.08 | 93 | 0.005 | 0.16 | 0.15 | 18 | −0.006 | 0.19 | 0.18 | 41 | 62 |
| | | $\beta^{(7)}$ | Large | −0.001 | 0.09 | 0.08 | 78 | −0.017 | 0.18 | 0.18 | 14 | 0.011 | 0.22 | 0.21 | 28 | 89 |
| | | | Medium | 0.000 | 0.08 | 0.08 | 95 | 0.002 | 0.15 | 0.15 | 21 | −0.007 | 0.18 | 0.18 | 43 | 97 |
| | 0.9 | $\beta^{(0)}$ | Large | 0.001 | 0.07 | 0.07 | 99 | 0.003 | 0.12 | 0.12 | 33 | −0.007 | 0.15 | 0.15 | 54 | 5 |
| | | | Medium | −0.002 | 0.07 | 0.07 | 94 | 0.003 | 0.10 | 0.10 | 45 | −0.004 | 0.13 | 0.13 | 68 | 4 |
| | | $\beta^{(4)}$ | Large | −0.002 | 0.08 | 0.07 | 93 | 0.005 | 0.12 | 0.11 | 33 | −0.007 | 0.16 | 0.15 | 58 | 78 |
| | | | Medium | 0.000 | 0.07 | 0.07 | 100 | 0.006 | 0.10 | 0.10 | 45 | −0.007 | 0.14 | 0.13 | 77 | 86 |
| | | $\beta^{(7)}$ | Large | −0.004 | 0.08 | 0.08 | 86 | −0.009 | 0.12 | 0.12 | 32 | 0.004 | 0.17 | 0.15 | 49 | 99 |
| | | | Medium | −0.001 | 0.08 | 0.08 | 100 | −0.003 | 0.10 | 0.10 | 47 | −0.003 | 0.14 | 0.14 | 72 | 100 |
| BIP + CPV | 0.6 | $\beta^{(0)}$ | Large | 0.000 | 0.08 | 0.07 | 82 | −0.002 | 0.14 | 0.13 | 24 | −0.007 | 0.19 | 0.18 | 34 | 6 |
| | | | Medium | −0.001 | 0.08 | 0.07 | 79 | −0.004 | 0.12 | 0.11 | 32 | 0.004 | 0.16 | 0.15 | 48 | 6 |
| | | $\beta^{(4)}$ | Large | 0.001 | 0.08 | 0.08 | 88 | −0.003 | 0.13 | 0.13 | 26 | 0.002 | 0.18 | 0.18 | 44 | 61 |
| | | | Medium | −0.004 | 0.08 | 0.08 | 88 | 0.004 | 0.11 | 0.11 | 35 | −0.005 | 0.15 | 0.16 | 59 | 74 |
| | | $\beta^{(7)}$ | Large | 0.003 | 0.08 | 0.08 | 88 | −0.001 | 0.13 | 0.13 | 26 | 0.007 | 0.18 | 0.18 | 41 | 96 |
| | | | Medium | −0.003 | 0.08 | 0.08 | 97 | −0.004 | 0.11 | 0.11 | 36 | 0.002 | 0.16 | 0.16 | 51 | 99 |
| | 0.9 | $\beta^{(0)}$ | Large | −0.002 | 0.07 | 0.07 | 91 | −0.002 | 0.10 | 0.10 | 47 | −0.006 | 0.15 | 0.14 | 57 | 7 |
| | | | Medium | −0.004 | 0.07 | 0.07 | 89 | 0.001 | 0.09 | 0.08 | 60 | −0.002 | 0.13 | 0.13 | 69 | 8 |
| | | $\beta^{(4)}$ | Large | −0.001 | 0.08 | 0.07 | 95 | 0.002 | 0.10 | 0.10 | 49 | −0.002 | 0.14 | 0.14 | 69 | 81 |
| | | | Medium | −0.005 | 0.08 | 0.07 | 95 | 0.005 | 0.08 | 0.08 | 66 | −0.004 | 0.13 | 0.13 | 84 | 86 |
| | | $\beta^{(7)}$ | Large | 0.002 | 0.08 | 0.08 | 95 | 0.000 | 0.09 | 0.10 | 53 | 0.006 | 0.14 | 0.15 | 67 | 100 |
| | | | Medium | −0.004 | 0.08 | 0.08 | 100 | 0.000 | 0.08 | 0.08 | 64 | 0.000 | 0.14 | 0.13 | 72 | 100 |

<p style="text-align:center">TABLE 1<br><i>(Continued)</i></p>

| Design | $\rho$ | $\beta$ | Missing | $\hat{\beta}_1$ | | | | $\hat{\beta}_2$ | | | | $\hat{\beta}_3$ | | | | |
|---|---|---|---|---|---|---|---|---|---|---|---|---|---|---|---|---|
| | | | | Bias | SD | SE | RE | Bias | SE | ASE | RE | Bias | SD | SE | RE | Power |
| CPV | | $\beta^{(0)}$ | Large | 0.015 | 0.10 | 0.09 | 51 | −0.023 | 0.26 | 0.24 | 7 | 0.031 | 0.31 | 0.29 | 12 | 8 |
| | | | Medium | 0.011 | 0.08 | 0.08 | 65 | −0.027 | 0.18 | 0.18 | 14 | 0.032 | 0.22 | 0.21 | 24 | 7 |
| | | $\beta^{(4)}$ | Large | 0.001 | 0.09 | 0.09 | 62 | −0.005 | 0.24 | 0.24 | 8 | −0.003 | 0.29 | 0.29 | 17 | 24 |
| | | | Medium | 0.001 | 0.09 | 0.08 | 69 | −0.012 | 0.18 | 0.18 | 13 | 0.008 | 0.22 | 0.21 | 29 | 47 |
| | | $\beta^{(7)}$ | Large | 0.002 | 0.10 | 0.09 | 67 | 0.001 | 0.23 | 0.24 | 8 | −0.002 | 0.28 | 0.29 | 17 | 70 |
| | | | Medium | −0.001 | 0.09 | 0.09 | 73 | 0.001 | 0.17 | 0.18 | 16 | −0.004 | 0.21 | 0.21 | 30 | 91 |

NOTE. $\beta^{(0)} = (−0.693, −1.109, 0)$; $\beta^{(4)} = (−0.849, −1.109, −0.4)$; $\beta^{(7)} = (−0.996, −1.109, −0.7)$. SE = Monte Carlo standard error, ASE = average of the bootstrap standard error from 50 bootstrap samples; RE = relative efficiency (ASE(gold standard)$^2$/ASE (missing)$^2$) × 100%; Power is for testing $H_0 : \beta_3 = 0$. "Large Missing" and "Medium Missing" patterns indicate the $IC$ size of 25% or 50% with additional 25% or 37.5% BIP data for designs with BIP, and include closeout $S^c(1)$ data from 25% or 50% uninfected placebo recipients for designs with CPV, respectively.

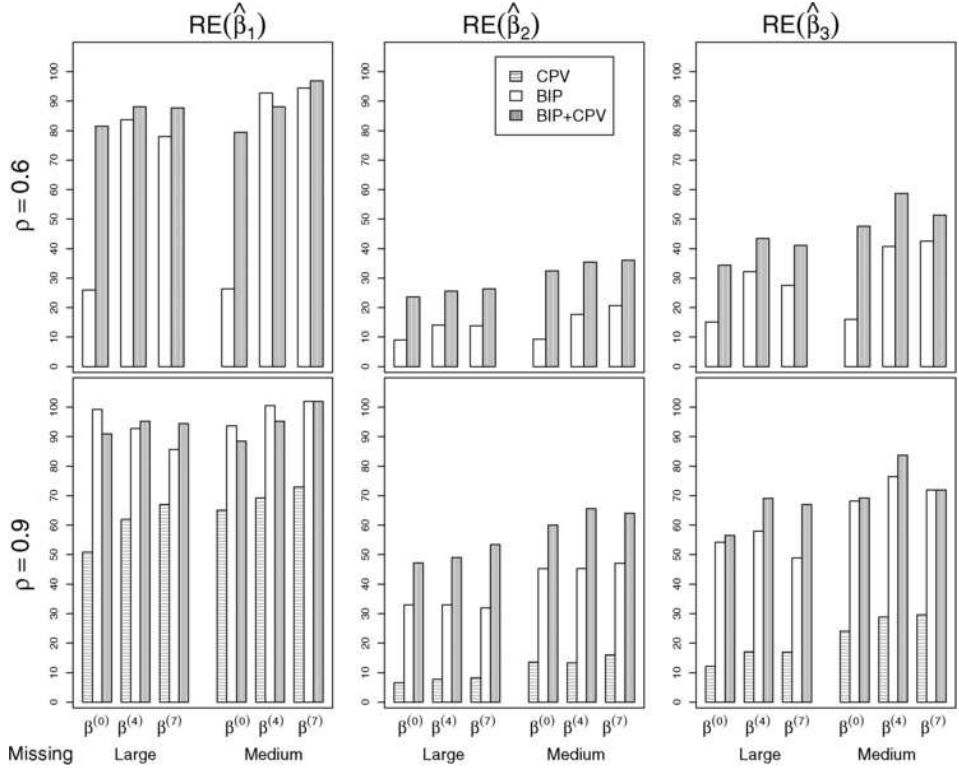

FIG. 3. *Relative efficiencies of parameter estimators. For designs with BIP, "Large Missing" and "Medium Missing" patterns indicate the IC size of 25% or 50% with additional 25% or 37.5% BIP data, respectively. For the design with CPV, "Large missing" and "Medium missing" patterns include closeout $S^c(1)$ data from 25% or 50% uninfected placebo recipients, respectively. True values of $\boldsymbol{\beta}$ ($\boldsymbol{\beta}^{(0)}, \boldsymbol{\beta}^{(4)}, \boldsymbol{\beta}^{(7)}$) are as specified in Tables 1 and 2.*

efficient because none of the infected placebo subjects have $S^c(1)$ measured. Figure 3 displays the relative efficiencies of the parameter estimators from the three designs with missing $S(1)$ with respect to the gold standard estimators. Overall the relative efficiency increases as the amount of measured immune responses increases. The relative efficiency of $\hat{\beta}_2$ is largely impacted by the amount of missing data, while that of $\hat{\beta}_1$ is less sensitive to the missing data pattern. These results confirm our design assumptions quite well.

Table 2 lists results from both the MEL approach and the ACEM algorithm under the BIP-alone design and the medium missing case (the $IC$ size of 50% with additional 37.5% first phase BIP data). It demonstrates that the performance of the ACEM method is very sensitive to the prediction accuracy of the baseline predictor. When the BIP is a fairly inaccurate predictor of $S(1)$ ($\rho = 0.6$), the ACEM method produces large biases and does

TABLE 2
*Comparison of results between the MEL and ACEM approaches for the BIP-alone design with the "Medium Missing" pattern (the IC size of 50% with additional 37.5% BIP data)*

| $\rho$ | $\beta$ | Method | $\hat{\beta}_1$ | | | | $\hat{\beta}_2$ | | | | $\hat{\beta}_3$ | | | | |
|---|---|---|---|---|---|---|---|---|---|---|---|---|---|---|---|
| | | | Bias | SE | ASE | RE | Bias | SE | ASE | RE | Bias | SE | ASE | RE | Power |
| 0.6 | $\beta^{(0)}$ | ACEM | $-0.139$ | 0.12 | 0.12 | 99 | 0.025 | 0.19 | 0.20 | 38 | $-0.238$ | 0.26 | 0.26 | 57 | 14 |
| | | MEL | 0.000 | 0.13 | 0.14 | 87 | $-0.002$ | 0.20 | 0.22 | 33 | 0.004 | 0.26 | 0.27 | 59 | 5 |
| | $\beta^{(4)}$ | ACEM | $-0.151$ | 0.13 | 0.13 | 100 | 0.019 | 0.19 | 0.20 | 37 | $-0.265$ | 0.26 | 0.26 | 64 | 72 |
| | | MEL | 0.004 | 0.14 | 0.14 | 88 | $-0.013$ | 0.21 | 0.21 | 31 | 0.013 | 0.26 | 0.27 | 60 | 34 |
| | $\beta^{(7)}$ | ACEM | $-0.162$ | 0.15 | 0.14 | 98 | 0.021 | 0.20 | 0.19 | 30 | $-0.292$ | 0.27 | 0.26 | 60 | 95 |
| | | MEL | 0.006 | 0.15 | 0.14 | 79 | $-0.008$ | 0.21 | 0.20 | 27 | 0.012 | 0.26 | 0.25 | 51 | 73 |
| 0.9 | $\beta^{(0)}$ | ACEM | $-0.043$ | 0.12 | 0.12 | 99 | 0.012 | 0.13 | 0.13 | 80 | $-0.064$ | 0.21 | 0.21 | 86 | 6 |
| | | MEL | $-0.002$ | 0.12 | 0.13 | 98 | $-0.004$ | 0.14 | 0.14 | 75 | 0.007 | 0.21 | 0.21 | 88 | 6 |
| | $\beta^{(4)}$ | ACEM | $-0.046$ | 0.13 | 0.13 | 99 | 0.008 | 0.13 | 0.13 | 74 | $-0.072$ | 0.22 | 0.21 | 89 | 58 |
| | | MEL | 0.000 | 0.13 | 0.13 | 97 | $-0.008$ | 0.14 | 0.14 | 69 | 0.008 | 0.21 | 0.21 | 87 | 45 |
| | $\beta^{(7)}$ | ACEM | $-0.054$ | 0.15 | 0.14 | 96 | 0.012 | 0.14 | 0.13 | 70 | $-0.094$ | 0.23 | 0.22 | 86 | 95 |
| | | MEL | 0.002 | 0.14 | 0.13 | 89 | $-0.004$ | 0.14 | 0.13 | 64 | 0.008 | 0.21 | 0.20 | 79 | 88 |

NOTE. $\beta^{(0)} = (-0.693, -1.109, 0)$; $\beta^{(4)} = (-0.849, -1.109, -0.4)$; $\beta^{(7)} = (-0.996, -1.109, -0.7)$. SE = Monte Carlo standard error, ASE = average of the bootstrap standard error from 50 bootstrap samples; RE = relative efficiency (ASE(gold standard)$^2$/ASE (missing)$^2$) × 100%; Power is for testing $H_0 : \beta_3 = 0$.

not control the type-I error rate; while if the calibration is reliable ($\rho = 0.9$), then the ACEM algorithm can generally estimate well. The MEL estimation outperforms the ACEM in most settings, with a slight loss of efficiency due to the grouping of the survival times.

**6. Discussion.** We have proposed a framework for assessing an immunological predictive surrogate in a vaccine trial with a time to event endpoint and case-cohort sampling of the immunological biomarker. While we have focused on the methods development for vaccine trials, the proposed principles are applicable for evaluating predictive surrogate endpoints in other biomedical applications.

We have discussed study designs and estimation procedures, and provided simulation results to demonstrate their validity and applicability under assumptions. We plan to apply the BIP-alone design to the three ongoing or pending HIV vaccine efficacy trials. As demonstrated by the simulation study, if good baseline irrelevant predictors exist, then a predictive surrogate can be evaluated effectively. The CPV-alone design is also a useful tool for the assessment that is complimentary to the BIP-alone design. If resources permit, the BIP + CPV design merits consideration because it improves accuracy and efficiency compared to the BIP-alone design if baseline predictors are not closely correlated with the potential predictive surrogate, or if A4 appears to be violated (i.e., the BIP affects clinical risk after controlling for the potential surrogate and first-phase baseline covariates).

For simplicity, we assumed equal drop-out and risk for each subject under assignment to vaccine or placebo over the time interval $[0, t_1]$ (assumption A3), and restricted the analysis to subjects at risk at the time the immune response is measured, $t_1$. To include all randomized subjects, A3 can be relaxed by postulating that the future immune response that will be measured at time $t_1$ impacts the risk of infection over $[0, t_1]$. With the DFT Cox model (1), the likelihood contribution of a subject with early infection during $[0, t_1]$ can be obtained as $\int \{1 - (1 - \lambda_{01})^{\exp\{V_i \beta_1 + s\beta_2 + V_i s\beta_3 + W_i' \boldsymbol{\beta}_4\}}\} \, dP(s)$, where $P(\cdot)$ is the marginal ($P(s)$) or conditional distribution of $S(1)$ ($P(s|B_i)$) if the BIP $B_i$ is measured. Another way to potentially weaken A3 would be to assume equal infection probabilities in $[0, t_1]$ for the vaccine and placebo groups, but not require that the vaccine has no effect for every individual.

A4 is a strong untestable assumption. Because we assume $B$ and $S(1)$ are correlated, A4 implies that the phase one covariates $W$ capture all the causes of $S(1)$ and the clinical endpoint [in the sense of Pearl (2000)]. Furthermore, it may be difficult to find a baseline covariate $B$ that is known to not affect clinical risk after accounting for $S(1)$. We suggest three potentially useful $B$'s for vaccine trials. First, a study that vaccinated 75 individuals simultaneously with hepatitis A and B vaccines showed a linear correlation of 0.85 among A- and B-specific antibody titers [Czeschinski, Binding and Witting (2000)]. Given there is little cross-reactivity among the hepatitis A and B proteins, $B$ = hepatitis A titer may be an excellent baseline predictor for $S(1)$ = hepatitis B titer that satisfies A4. For HIV vaccine trials, two available scalar $B$'s may plausibly satisfy A4. First, Follmann (2006) considered as $B$ the antibody titer to a rabies glycoprotein vaccine. Because rabies is not acquired sexually, it is plausible that anti-rabies antibodies are independent of risk of HIV infection given $S(1)$. Second, in the ongoing HIV vaccine efficacy trials, a current leading candidate $B$ is the titer of antibodies that neutralize the Adenovirus serotype 5 vector that carries the HIV genes in the vaccine. This $B$ has been shown to inversely correlate with the $S(1)$ of primary interest (T cell response levels measured by ELISpot) [Catanzaro et al. (2006)], and since Adenovirus 5 is a respiratory infection virus, A4 may plausibly hold.

In general, though, it is desirable to relax A4, and fortunately this can be done by including $B$ as a component of $W$ in the Cox model (1) and estimating its coefficient (as suggested by the Associate Editor). This extra coefficient for $B$ is identified by the data from vaccine recipients with $B$ measured. Based on the argument given by Follmann (2006) and Gilbert and Hudgens (2006) for the setting of the BIP-alone design and a dichotomous clinical endpoint, we conjecture that the estimand $VE(s_1)$ will be identified from the observed data as long as at least one of the interaction terms of $B$ with $V$ or $W$ with $V$ is omitted from the Cox model.

Our approach specified a Cox regression with $S_i(1)$ and $S_i(1)V$ as covariates. Another approach is to assume that the immune response is measured with some "error," $S_i(1) = S_i^{\text{true}}(1) + e_{i1}$ and $S_i^c(1) = S_i^{\text{true}}(1) + e_{i2}$ (as is done in A5), but then to use the true immune responses $S_i^{\text{true}}(1)$ and $S_i^{\text{true}}(1)V$ as covariates in the Cox model. To proceed with this model, one could obtain replicates of $S_i(1)$ and $S_i^c(1)$, say, $S_{i1}(1), S_{i2}(1)$ and $S_{i1}^c(1), S_{i2}^c(1)$, and assume that the $e_i$s followed a Gaussian distribution with mean 0 and unknown variance $\tau^2$. Then a more complicated likelihood could be written by integrating $S_i^{\text{true}}(1)$ over the distribution of $S_i^{\text{true}}(1)|S_{i1}(1), S_{i2}(1)$, $S_i^{\text{true}}(1)|S_{i1}^c(1), S_{i2}^c(1)$, or $S_i^{\text{true}}(1)|B_i$ as appropriate.

We have presented estimated likelihood based methods to accommodate missing data in case-cohort designs, as well as a regression calibration based double-semiparametric EM algorithm that has reasonable performance when the regression calibration is reliable and the event is rare. This approximate algorithm enjoys the convenience of regression calibration to incorporate auxiliary information, and has faster and easier implementation for the continuous failure time model. Alternative estimation methods such as multiple imputation may also be useful, provided the posterior distribution can be properly specified. In addition, a full likelihood approach that maximized over $(\boldsymbol{\beta}, \boldsymbol{\lambda}_0)$ and the parameters of $p(s|w)$ and $p(s|b,w)$ all at once could be used. While the full likelihood should be more efficient if the entire joint model is correctly specified, MEL is simpler to implement and may be more robust to joint model mis-specification.

## APPENDIX: INTEGRAL CALCULATION IN LIKELIHOOD (??)

For discrete $S(1)$ and $B$, the integrations can be replaced by finite summations. When $S(1)$ is continuous, the integrations can be made easier by positing parametric models. Assume $S(1) \sim \mathrm{N}(\mu(\cdot), \sigma(\cdot)^2)$, where $\mu(\cdot), \sigma(\cdot)^2$ represent the first two moments of $p(s)$ or $p(s|b)$. Then for a given function $g(s)$ of $s$, $\int g(s)p(\cdot)\,ds = \int g(\mu(\cdot) + \sigma(\cdot)u)\phi(u)\,du$, where $p(\cdot)$ denotes $p(s)$ or $p(s|b)$ and $\phi(u)$ is the standard normal density function. Because the integrand $g(s)$ in (4) is a smooth function of $s$, numerical methods such as Gaussian quadrature can be applied to evaluate the integration. Based on our experience, only a small number (around 15) of evaluations is needed to get stable quadrature results.

When $B$ has discrete values $b_l, l = 1, \ldots, L$, an alternative way to integrate over $s$ is through the nonparametric representation of $p(s)$ and $p(s|b)$. The integrals $\int g(s)p(\cdot)\,ds$ can be evaluated nonparametrically by

$$
\int g(s)p(s)\,ds \approx p_{11} \frac{1}{\sum_{i \in IC_V} \delta_i} \sum_{i \in IC_V} \delta_i g(S_i(1))
$$
$$
+ p_{10} \frac{1}{\sum_{i \in IC_V} (1-\delta_i)} \sum_{i \in IC_V} (1-\delta_i) g(S_i(1)),
$$
$$
\int g(s)p(s|b_l)\,ds \approx p_{11} \frac{1}{\sum_{i \in IC_V, B_i=b_l} \delta_i} \sum_{i \in IC_V, B_i=b_l} \delta_i g(S_i(1))
$$
$$
+ p_{10} \frac{1}{\sum_{i \in IC_V, B_i=b_l} (1-\delta_i)} \sum_{i \in IC_V, B_i=b_l} (1-\delta_i) g(S_i(1)).
$$

**Acknowledgments.** The authors thank Huayun Chen for providing his fortran code for the EM algorithm of Chen (2002), and the Editor, Associate Editor and Referees for helpful comments.

## REFERENCES

Borgan, O., Langholz, B., Samuelsen, O., Goldstein, L. and Pogod, J. (2000). Exposure stratified case-cohort designs. *Lifetime Data Anal.* **6** 39–58. MR1767493

Catanzaro, A., Koup, R., Roederer, M. et al. (2006). Safety and immunogenicity evaluation of a multiclade HIV-1 candidate vaccine delivered by a replication-defective recombinant adenovirus vector. *J. Infectious Diseases* **194** 1638–1649.

Chan, I., Wang, W. and Heyse, J. (2003). *Vaccine Clinical Trials in Encyclopedia of Biopharmaceutical Statistics*, 2nd ed. Dekker, New York.

CHEN, H. Y. (2002). Double-semiparametric method for missing covariates in Cox regression models. *J. Amer. Statist. Assoc.* **97** 565–576. MR1941473

CHEN, H. Y. and LITTLE, R. J. A. (1999). Proportional hazards regression with missing covariates. *J. Amer. Statist. Assoc.* **94** 896–908. MR1723311

COX, D. R. (1972). Regression models and life-tables (with discussion). *J. Roy. Statist. Soc. Ser. B* **34** 187–220. MR0341758

CZESCHINSKI, P., BINDING, N. and WITTING, U. (2000). Hepatitis A and hepatitis B vaccinations: immunogenicity of combined vaccine and of simultaneously or separately applied single vaccines. *Vaccine* **18** 1074–1080.

DEGRUTTOLA, V. G., CLAX, P., DEMETS, D. L., DOWNING, G. J., ELLENBERG, S. S., FRIEDMAN, L., GAIL, M. H., PRENTICE, R., WITTES, J. and ZEGER, S. L. (2002). Considerations in the evaluation of surrogate endpoints in clinical trials. *Controlled Clinical Trials* **22** 485–502.

FOLLMANN, D. (2006). Augmented designs to assess immune response in vaccine trials. *Biometrics* **62** 1161–1169. MR2307441

FRANGAKIS, C. and RUBIN, D. (2002). Principal stratification in causal inference. *Biometrics* **58** 21–29. MR1891039

GILBERT, P. B. and HUDGENS, M. G. (2006). Evaluating causal effect predictiveness of candidate surrogate endpoints. Working Paper 291, UW Biostatistics Working Paper Series.

GILBERT, P. B., BOSCH, R. and HUDGENS, M. G. (2003). Sensitivity analysis for the assessment of causal vaccine effects on viral loads in HIV vaccine trials. *Biometrics* **59** 531–541. MR2004258

GILBERT, P. B., PETERSON, M., FOLLMANN, D., HUDGENS, M., FRANCIS, D., GURWITH, M., HEYWARD, W., JOBES, D., POPOVIC, V., SELF, S., SINANGIL, F., BURKE, D. and BERMAN, P. (2005). Correlation between immunologic responses to a recombinant glycoprotein 120 vaccine and incidence of HIV-1 infection in a phase 3 HIV-1 preventive vaccine trial. *J. Infectious Diseases* **191** 666–677.

HALLORAN, M. E. (1998). Vaccine studies. In *Encyclopedia of Biostatistics* (P. Armitage and T. Colton, eds.) 4687–4694. Wiley, New York.

HERRING, A. H. and IBRAHIM, J. G. (2001). Likelihood-based methods for missing covariates in the Cox proportional hazards model. *J. Amer. Statist. Assoc.* **96** 292–302. MR1952739

HOLLAND, P. (1986). Statistics and causal inference (with discussion). *J. Amer. Statist. Assoc.* **81** 945–970. MR0867618

KULICH, M. and LIN, D. Y. (2004). Improving the efficiency of relative-risk estimation in case-cohort studies. *J. Amer. Statist. Assoc.* **99** 832–844. MR2090916

LIN, D., FLEMING, T. R. and DE GRUTTOLA, V. (1997). Estimating the proportion of treatment effect explained by a surrogate markers. *Statistics in Medicine* **16** 1515–1527.

LIN, D. Y. and YING, Z. (1993). Cox regression with incomplete covariate measurements. *J. Amer. Statist. Assoc.* **88** 1341–1349. MR1245368

LITTLE, R. A. and RUBIN, D. B. (2002). *Statistical Analysis of Missing Data*, 2nd ed. Wiley, Hoboken, NJ. MR1925014

MEHROTRA, D. V., LI, X. and GILBERT, P. B. (2006). A comparison of eight methods for the dual-endpoint evaluation of efficacy in a proof-of-concept HIV vaccine trial. *Biometrics* **62** 893–900. MR2247219

MOLENBERGHS, G., BUYSE, M., GEYS, H., RENARD, D., BURZYKOWSKI, T. and ALONSO, A. (2002). Statistical challenges in the evaluation of surrogate endpoints in randomized trials. *Controlled Clinical Trials* **23** 607–625.

PAIK, M. C. and TSAI, W. Y. (1997). On using the Cox proportional hazards model with missing covariates. *Biometrika* **84** 579–593. MR1603989

PEARL, J. (2000). *Causality: Models, Reasoning, and Inference.* Cambridge Univ. Press. MR1744773

PEPE, M. and FLEMING, T. (1991). A nonparametric method for dealing with mismeasured covariate data. *J. Amer. Statist. Assoc.* **86** 108–113. MR1137103

PRENTICE, R. L. (1982). Covariate measurement errors and parameter estimation in a failure time regression model. *Biometrika* **69** 331–342. MR0671971

PRENTICE, R. L. (1986). A case-cohort design for epidemiologic cohort studies and disease prevention trials. *Biometrika* **73** 1–11.

PRENTICE, R. L. (1989). Surrogate endpoints in clinical trials: Definition and operational criteria. *Statistics in Medicine* **8** 431–440.

ROBINS, J. M., ROTNITZKY, A. and ZHAO, L. P. (1994). Estimation of regression coefficients when some regressors are not always observed. *J. Amer. Statist. Assoc.* **89** 846–866. MR1294730

RUBIN, D. B. (2005). Causal inference using potential outcomes: Design, modeling, decisions. *J. Amer. Statist. Assoc.* **100** 322–331. MR2166071

SCHEIKE, T. H. and MARTINUSSEN, T. (2004). Maximum likelihood estimation for Cox's regression model under case-cohort sampling. *Scand. J. Statist.* **31** 283–293. MR2066254

WANG, C. Y., HSU, L., FENG, Z. D. and PRENTICE, R. L. (1997). Regression calibration in failure time regression. *Biometrics* **53** 131–145. MR1450183

WEIR, C. and WALLEY, R. (2006). Statistical evaluation of biomarkers as surrogate endpoints: A literature review. *Statistics in Medicine* **25** 183–203. MR2222082

ZHOU, H. B. and PEPE, M. S. (1995). Auxiliary covariate data in failure time regression. *Biometrika* **82** 139–149. MR1332845

L. QIN
P. B. GILBERT
VACCINE AND INFECTIOUS DISEASE INSTITUTE
FRED HUTCHINSON CANCER RESEARCH CENTER
DEPARTMENT OF BIOSTATISTICS
UNIVERSITY OF WASHINGTON
1100 FAIRVIEW AVENUE NORTH, LE-400
SEATTLE, WASHINGTON 98109
USA
E-MAIL: lqin@scharp.org
            pgilbert@scharp.org

D. FOLLMANN
NATIONAL INSTITUTE OF ALLERGY
   AND INFECTIOUS DISEASES
6700B ROCKLEDGE DRIVE MSC 7609
BETHESDA, MARYLAND 20892
USA
E-MAIL: dfollmann@niaid.nih.gov

D. LI
SCHOOL OF MATHEMATICAL SCIENCE
PEKING UNIVERSITY
BEIJING 100871
P.R. CHINA
E-MAIL: ldf@math.pku.edu.cn